\pdfoutput=1
\documentclass[aps,prl,twocolumn,superscriptaddress]{revtex4-1}
\usepackage{graphicx}
\usepackage[utf8]{inputenc}

\begin{document}
\title{Ground-state spin blockade in a single-molecule junction}

\author{J. de Bruijckere}
\affiliation{Kavli Institute of Nanoscience, Delft University of Technology, 2628 CJ Delft, The Netherlands}
\author{P. Gehring} 
\affiliation{Kavli Institute of Nanoscience, Delft University of Technology, 2628 CJ Delft, The Netherlands}
\author{M. Palacios-Corella}
\affiliation{Instituto de Ciencia Molecular (ICMol), Universidad de Valencia, Catedr{\'a}tico Jos{\'e} Beltr{\'a}n 2, Paterna, 46980, Spain}
\author{M. Clemente-Le{\'o}n}
\affiliation{Instituto de Ciencia Molecular (ICMol), Universidad de Valencia, Catedr{\'a}tico Jos{\'e} Beltr{\'a}n 2, Paterna, 46980, Spain}
\author{E. Coronado}
\affiliation{Instituto de Ciencia Molecular (ICMol), Universidad de Valencia, Catedr{\'a}tico Jos{\'e} Beltr{\'a}n 2, Paterna, 46980, Spain}
\author{J. Paaske}
\affiliation{Niels Bohr Institute, University of Copenhagen, DK-2100 Copenhagen, Denmark}
\affiliation{Center for Quantum Devices, Niels Bohr Institute, University of Copenhagen, DK-2100 Copenhagen, Denmark}
\author{P. Hedeg{\aa}rd}
\affiliation{Niels Bohr Institute, University of Copenhagen, DK-2100 Copenhagen, Denmark}
\author{H.S.J. van der Zant}
\email[E-mail: ]{H.S.J.vanderZant@tudelft.nl}
\affiliation{Kavli Institute of Nanoscience, Delft University of Technology, 2628 CJ Delft, The Netherlands}

\date{\today}

\begin{abstract}
It is known that the quantum-mechanical ground state of a nano-scale junction has a significant impact on its electrical transport properties. This becomes particularly important in transistors consisting of a single molecule. Due to strong electron-electron interactions and the possibility to access ground states with high spins, these systems are eligible hosts of a current-blockade phenomenon called ground-state spin blockade. This effect arises from the inability of a charge carrier to account for the spin difference required to enter the junction, as that process would violate the spin selection rules. Here, we present a direct experimental demonstration of ground-state spin blockade in a high-spin single-molecule transistor. The measured transport characteristics of this device exhibit a complete suppression of resonant transport due to a ground-state spin difference of 3/2 between subsequent charge states. Strikingly, the blockade can be reversibly lifted by driving the system through a magnetic ground-state transition in one charge state, using the tunability offered by both magnetic and electric fields.
\end{abstract}

\maketitle

%% INTRODUCTION
Blockade mechanisms in charge transport involve various physical phenomena: The current in double quantum dots can be inhibited by the Pauli exclusion principle \cite{koppens2005control, johnson2005singlet, liu2008pauli}, vibrational transitions in molecular junctions and quantum dots can be suppressed by Franck-Condon blockade \cite{koch2005franck,leturcq2009franck,burzuri2014franck}, and in junctions with superconducting electrodes there may be no low-energy transport as a result of the superconducting energy gap \cite{winkelmann09,deng2016majorana}. These blockade mechanisms all occur in combination with Coulomb blockade, which results from the energy level spacing $\Delta E$ and the energy costs of charging a weakly-coupled nano-object, i.e., the charging energy $U$. Coulomb blockade can be overcome as soon as the energy of an electron in one electrode exceeds the addition energy $E_{\text{add}}=\Delta E+U$. Then, electrons can sequentially travel from one electrode to the other via the nano-object. In every sequential electron tunneling (SET) event, the charge of the object changes by an elementary charge and the spin typically by 1/2, due to the added or removed electron. 

Here, we study an exceptionally clear manifestation of the blockade phenomenon that occurs when the ground-state spin of subsequent charge states differs by more than 1/2. SET transitions between these ground states are forbidden by the spin selection rules and Coulomb blockade peaks are suppressed \cite{weinmann1995spin,heersche2006electron,gaudenzi2017transport}. We refer to this effect as ground-state spin blockade (GSSB). One of the necessary requirements for GSSB is a high-spin ground state, i.e., $S>1$, in one of the charge states. This requirement is hard to obtain for top-down quantum dots and in the few earlier demonstrations of GSSB, neither the full suppression of the Coulomb blockade peaks nor control over the blockade was achieved \cite{rokhinson2001spin,huttel2003spin}. For molecules, high-spin ground states can be tailored by chemical design, making them promising candidates for observing GSSB. In this transport study, we provide experimental evidence for complete GSSB in a high-spin single-molecule junction. With an external magnetic field, the blockade can be reversibly lifted by driving the molecule in one charge state through a magnetic ground-state transition. The presence of GSSB puts constraints on the allowed transitions and is used as a diagnostic tool to determine the ground state, and excited states of the molecule.

%% MEASUREMENT DESCRIPTION
The measurements were carried out with the device sketched in Fig. 1a. A single molecule is embedded in a circuit with two gold electrodes in which DC current ($I$) is measured as a function of the applied voltage difference between the electrodes (the bias voltage $V$) and the voltage applied to a capacitively coupled gate electrode ($V_{g}$). The junction is formed by room-temperature electromigration \cite{park99} and self-breaking \cite{oneill07} of a gold nanowire. A dilute solution of the molecules is dropcasted on a chip with 24 electromigrated junctions on which, after pumping away the solution and cooling down the sample, three junctions show Coulomb blockade with addition energies in the typical regime of molecular junctions, i.e., $E_{\text{add}}>100$ meV. The tunnel couplings and energy level alignments cannot be controlled by this technique and depend on the way the molecule is trapped inside the junction. One of the three samples shows the right combination of a small tunnel coupling, which allows for high-resolution spectroscopy, and a level alignment close to the Fermi level of the electrodes, such that the molecule can be charged within the accessible gate voltage range. All measurements shown here are of this particular junction and are taken at $T \approx$ 40 mK in a dilution refrigerator. 

\begin{figure}
\includegraphics{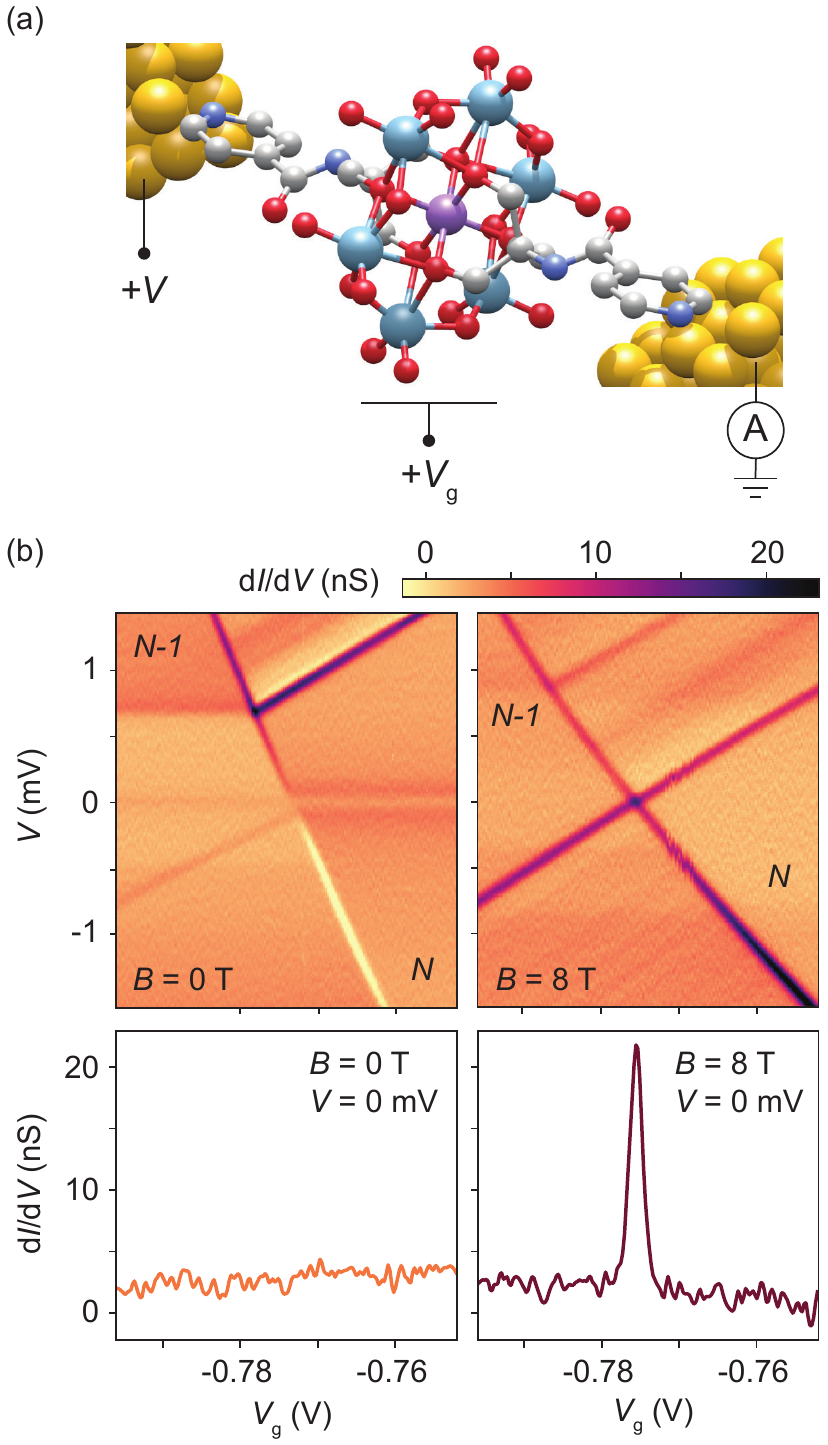}
\caption{(a) Sketch of the molecule, the gold electrodes and the measurement circuit. The molecule consists of a Mn(III) center (purple) surrounded by six MoO$_{6}$ octahedra, which connect on both sides to pyridine-based ligands. Electric current ($I$) through the molecule is recorded as a function of bias voltage ($V$) and gate voltage ($V_{\text{g}}$). (b), (top) d$I$/d$V$ maps at zero magnetic field (left) and at 8 T (right). At 0 T, no SET lines corresponding to transitions between the ground states of $N$ and $N-1$ are present, as a result of ground-state spin blockade. At 8 T, these lines do appear and the blockade is lifted. (bottom) d$I$/d$V$ traces at zero bias, showing a charge degeneracy peak at 8 T and its complete suppression at 0 T.}
\label{figure1}
\end{figure}

%% MOLECULE
The molecule in the junction is a pyridine-functionalized Mn(III) Anderson polyoxometalate (POM) \cite{allain2008hybrid}, which is sketched in Fig. 1a. It consists of a molecular metal oxide formed by a central Mn(III)-ion, surrounded by six edge-sharing octahedra, which confer robustness to the POM and magnetic isolation. This type of molecule is axially functionalized on both sides by organic molecules linked directly to the Mn(III)-ion through alcoxo bridges. Magnetic characterization of these compounds in the solid state shows that the Mn(III) has a total spin $S=2$ and a g-factor close to 2 \cite{abherve2015bimetallic} (see Supplemental Material for more details on the synthesis of the molecule).

%% DESCRIPTION FIGURE 1
In Fig. 1b, the two top panels show differential conductance (d$I$/d$V$) maps of the device as a function of $V$ and $V_{\text{g}}$, at zero magnetic field and at 8 T. For clarity, we first discuss the map at 8 T, which looks like a common d$I$/d$V$ map of a Coulomb-blockaded system containing a single object, here, a molecule. The regions labeled $N$ and $N-1$ are regions in which the charge of the molecule is fixed and SET processes are suppressed. Inside these regions, transport is governed by co-tunneling (COT) processes, which appear as horizontal lines. The two slanted lines that form the cross-like shape are the edges of the Coulomb diamonds and their point of incidence at zero bias is the charge degeneracy point. The bottom right panel shows that this point appears as a peak in the linear-conductance gate trace, i.e., the d$I$/d$V|_{V=0}$ as a function of $V_{\text{g}}$. Looking for peaks in these traces is typically the first step to check whether a molecule is present in the junction. In the top and bottom regions of the d$I$/d$V$ map, the charge of the molecule can fluctuate, allowing for SET processes to occur. The top SET region contains slanted lines starting from the left Coulomb edge at $V=0.9$ mV and, less intense, at $V=0.3$ mV, both moving towards the top right. These lines correspond to SET processes involving excited states.

Remarkably, in the map at 0 T, the charge degeneracy point is absent: slanted lines are present, but they do not cross at zero bias. At any gate voltage, the linear conductance is suppressed and only SET lines at finite voltages are present. This shows that transitions between the ground state of $N$ and the ground state of $N-1$ are blocked. The complete suppression can be seen more clearly in the zero-bias trace presented in the corresponding bottom panel, which in contrast to the 8 T trace shows no peak, or any other molecular signature. All the slanted lines in the map appear at finite bias voltages and mark transitions involving at least one excited state: The SET line starting at $V=0.7$ mV coincides with a COT line at that same bias voltage, which implies that their excitation energies are equal, and involve the same excited state. Similarly, the COT lines at $V=\pm 0.1$ mV in $N$ connect to two SET lines starting at the same bias voltages. Another faint COT line appears at $V=-0.5$ mV in $N-1$, closer to zero bias than its counterpart at positive bias. This asymmetry suggests that the corresponding excitation energy is influenced by the bias voltage, as will be discussed below. The region $N-1$ also contains a zero-bias line that resembles a Kondo resonance \cite{liang2002kondo}, resulting from a degenerate ground state.

\begin{figure}
\includegraphics{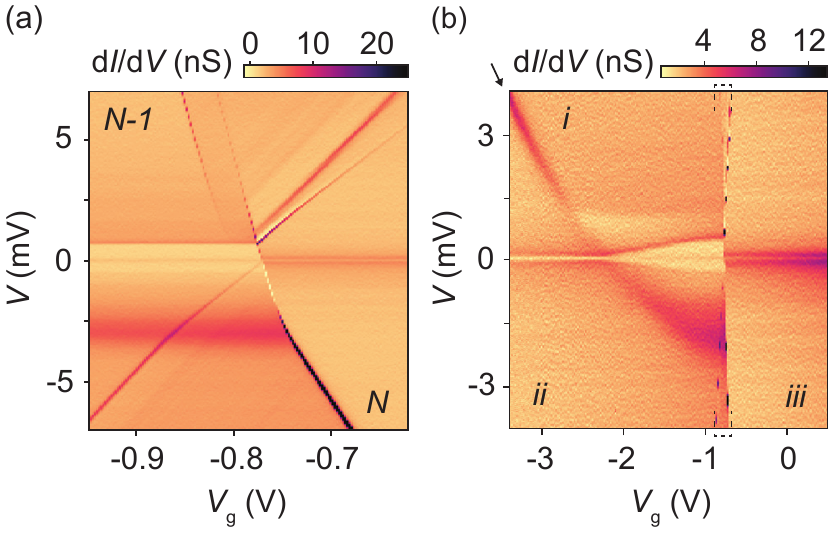}
\caption{d$I$/d$V$ maps at 0 T in a wide bias and gate voltage range. (a), d$I$/d$V$ map showing the same features as in Fig. 1b, along with a ground state transition line at $V=-3$ mV and additional SET excitation lines with different slopes. The SET lines at negative bias change slope at the ground state transition. (b), d$I$/d$V$ map in a larger gate window, where the region marked by dashed brackets is the region shown in (a). The dark parabolic feature indicated by the arrow is the ground state transition line in (a). Three different regions with different ground states are separated by the discontinuity and the parabola, labeled $i$, $ii$ and $iii$. The excitation energy of the COT excitations in $i$ are tunable by the gate voltage.}
\label{figure2}
\end{figure}

\begin{figure}
\includegraphics{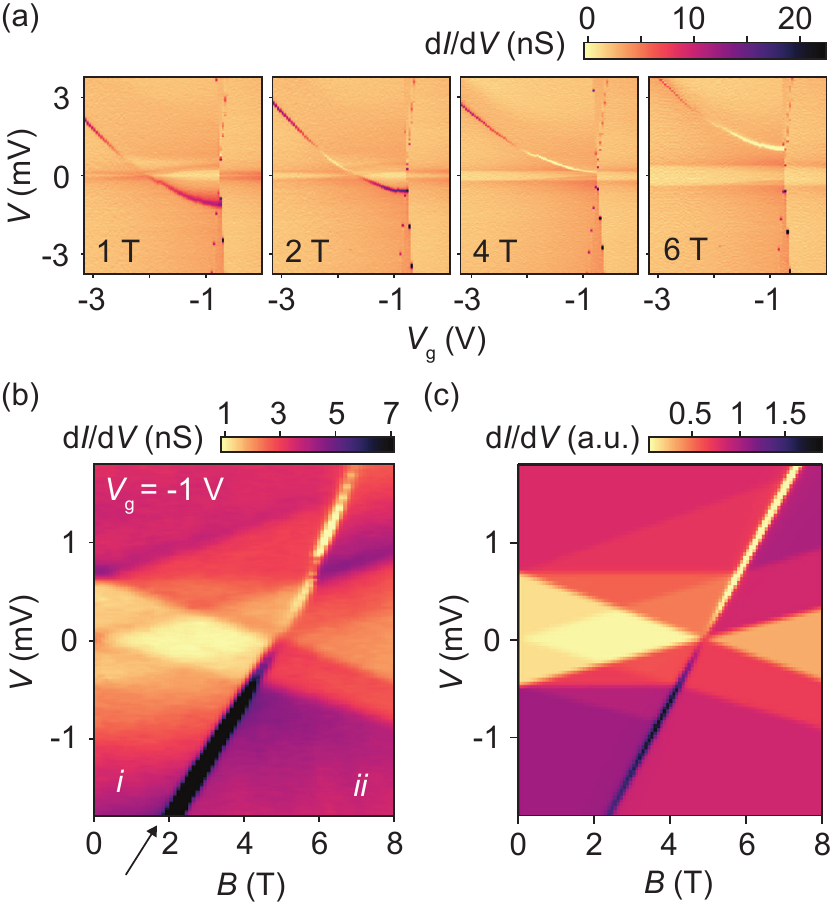}
\caption{(a) d$I$/d$V$ maps of the same region as Fig. 2b, at different magnetic fields. The parabolic line separating the two spin ground states in $N-1$ moves upwards by increasing the magnetic field as the high-spin state becomes energetically more favorable. (b), Magnetic field dependence of the d$I$/d$V$ spectrum at $V_{\text{g}} = -1.0$ V. One line originating from the excited multiplet crosses zero bias at $B = 5$ T and becomes the new ground state above this field. The two ground states are separated by the slanted line indicated by the arrow. (c), Simulated d$I$/d$V$ spectra for a system of a spin-1/2, tunnel coupled to two reservoirs and exchange coupled to a spin-3/2, with a $V$-dependent exchange coupling.}
\label{figure3}
\end{figure}

%% DESCRIPTION FIGURE 2
The striking difference between the maps at 0 T and at 8 T in Fig. 1b is the presence of the charge degeneracy point; the complete suppression of this point at 0 T is completely lifted at 8 T. To study the blockade mechanism in more detail, we show two d$I$/d$V$ maps in a larger gate and bias-voltage range in Fig. 2, both recorded at 0 T. In Fig. 2a, the same features as in Fig. 1b can be observed, along with a broad horizontal line at $V=-3$ mV and additional SET excitation lines at higher bias voltages. The Coulomb edges at negative bias change slope at the two coincidences with the broad horizontal line. This feature shows striking similarities with the simulated transport data presented in \cite{stevanato2012finite}, where a singlet-triplet ground-state transition line in a double-quantum-dot model is studied. The energies of the ground states separated by this line have a different dependence on bias voltage, causing their corresponding Coulomb edges to have different slopes.

The gate and bias-voltage dependence of the ground-state transition line is shown in Fig. 2b where the transport data is presented for an even wider gate-voltage range. The dashed brackets in this figure mark the gate-voltage range of Fig. 2a. Figure 2b reveals that the ground-state transition line, indicated by the arrow, extends towards positive bias voltages and has an anomalous parabolic shape; it can be well approximated by a function quadratic in $V_{\text{g}}$ and linear in $V$ (see Supplemental Material for a more detailed analysis of the ground-state transition line). Three regions with different ground states can be distinguished, separated by the parabola and the Coulomb edges. We label the regions $i$, $ii$ and $iii$, where $i$ and $ii$ belong to the charge state $N-1$, and $iii$ to the charge state $N$. In both $ii$ and $iii$, a pair of COT lines appears at $V=\pm 0.1$ mV, symmetrically positioned around zero bias. In $i$, a pair of gate-voltage dependent COT lines is present, starting at $V=\pm 0.1$ mV around $V_{\text{g}}=-2$ V, moving away from zero bias as $V_{\text{g}}$ is increased. The excitation energy of this COT line thus depends on the gate voltage.

%% DESCRIPTION FIGURE 3
Next, we investigate the magnetic field dependence of the parabolic ground-state transition line. Figure 3a shows four maps in the same bias and gate voltage range as in Fig. 2b, at different magnetic fields. By increasing the magnetic field, the parabola moves towards higher bias voltages, and the d$I$/d$V$ along the line changes in magnitude and sign. For an increasing part of the line, the d$I$/d$V$ turns negative, i.e., the current goes down by increasing the bias voltage at the transition from $ii$ to $i$.

The fact that the ground-state transition line moves upwards with magnetic field implies that the total spin of the ground state in region $ii$ is larger than in region $i$; upon increasing the magnetic field, the ground state with higher spin becomes energetically more favorable and the transition occurs at higher bias voltages. Above 4 T, region $ii$ moves across zero bias at the charge degeneracy point, which lifts the GSSB at this point. The parabola thus marks the transition from a spin-blockaded region ($i$), to a region where the blockade is lifted ($ii$).

To identify the spin states in $N-1$ we focus on the magnetic-field dependence of the d$I$/d$V$ spectrum at a fixed gate voltage $V_{\text{g}}=-1$ V (see Fig. 3b). The Kondo-like peak at zero bias splits linearly in two in a magnetic field, which verifies the presence of a degenerate ground state at 0 T. The COT excitations at $V=+0.7$ mV and $V=-0.4$ mV split in three and appear asymmetric in position and intensity. The fact that these excitations split in three implies that the corresponding excited state is a spin multiplet with a larger spin than the ground state; this can be deduced from the spin selection rules for COT processes \cite{gaudenzi2017transport}. Moreover, the spin selection rules impose additional constraints which lead to the conclusion that the spin difference between the excited state and the ground state is 1.

At about 5 T, one excitation from the excited spin multiplet crosses zero bias, which at that point becomes the new spin ground state. The two regions with different ground states are separated by the slanted line indicated by the arrow. These regions correspond to $i$ and $ii$ in Fig. 2b, labeled accordingly in Fig 3b. The finite slope of the ground-state transition line results from the influence of the bias voltage on the associated excitation energy. This is also reflected by the asymmetry in bias voltage at which the multiplet excitations appear.

A model explaining the main features of the experimental data can now be constructed. The observed spin excitations occur at relatively low bias voltages ($\sim$ 1 mV), which suggests that they are not related to spin reconfigurations of the Mn center itself, as for spin-crossover molecules \cite{miyamachi2012robust, meded2011electrical}; the energies of these transitions are typically orders of magnitude larger. Rather, we propose a model in which we invoke a weakly-coupled spin, exchange coupled to the high-spin center. This spin is possibly residing on the ligands of the molecule, as in other molecular systems \cite{thiele2014electrically,osorio2009electrical} and we will refer to it as the ligand spin. The low-energy COT excitations in this model are transitions in which the ligand spin is flipped with respect to the spin of the Mn center. 

We simulate the spin-excitation spectra of this system by a tunneling model based on \cite{ternes2015spin}. The result of the simulation is presented in Fig. 3c. The model includes COT processes of second order and solves the rate equations for a spin system consisting of a ligand spin $S=1/2$, tunnel coupled to two reservoirs and exchange coupled to a high-spin ($S=3/2$) center. The exchange coupling in $N-1$ is anti-ferromagnetic, which at low magnetic fields results in a low-spin ground state ($S=1$) and a high-spin excited state ($S=2$). We add a linear $V$ term in the expression of the exchange coupling as in \cite{stevanato2012finite}, which accounts for the asymmetric positions of the COT excitations and the appearance of the slanted ground-state transition line.

\begin{figure}
\includegraphics{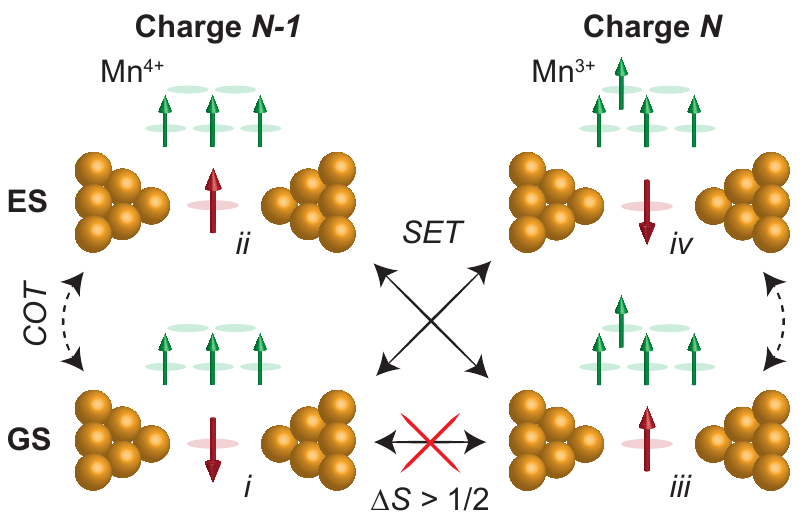}
\caption{Diagram showing the four states of different charge and spin, where the red and green arrows correspond to the ligand spin and the spins in the Mn center, respectively. The observed SET and COT transitions are represented by solid and dashed arrows, respectively. The red cross indicates the blocked SET transition. For charge state $N-1$, $S=1$ ($i$) and $S=2$ ($ii$), and for charge state $N$, $S=5/2$ ($iii$) and $S=3/2$ ($iv$).}
\label{figure4}
\end{figure}

%% DESCRIPTION FIGURE 4
The ground state (GS) and excited state (ES) in the simulation of $N-1$ are sketched on the left-hand side in Fig. 4, where red arrows represent the ligand spins, green arrows the spins of the Mn center and yellow spheres the gold atoms of the electrodes. The right-hand side of Fig. 4 shows the GS and ES in $N$. The observed SET and COT transitions between the states are indicated by the solid and dashed black arrows, respectively. In the transition from charge state $N-1$ to $N$, the added electron is likely to reside on the metal center, changing the oxidation state from Mn$^{4+}$ to Mn$^{3+}$, as illustrated in Fig. 4. This charge transition of the Mn center occurs in solutions of this compound as shown by cyclic voltammetry measurements \cite{allain2008hybrid}. The spin of the Mn center in $N$ would then match the spin measured in crystals of this type of compound \cite{abherve2015bimetallic}, namely $S=2$.

For GSSB to occur in this system, the ground-state spin should change by more than 1/2 upon charging. This can happen if the added charge not only contributes with its intrinsic spin, but its presence also changes the sign of the exchange coupling from anti-ferromagnetic to ferromagnetic, further increasing the total spin by 1. Such a transition of the exchange coupling is observed, and it is marked by the parabolic ground-state transition line in $N-1$. The gate and bias-voltage dependence of this line shows that the exchange coupling is strongly influenced by the electric field, as was observed in other molecular systems \cite{osorio2009electrical,roch2008quantum}.   

We speculate that the observed tunability of the exchange coupling between the ligand spin and the spin of the Mn center follows from the influence of the gate and bias voltage on the spatial distribution of the ligand spin, which appears to be different for the two charge states: In $N$, the ligand spin could be delocalized over the $e_{g}$ orbitals of the Mn$^{3+}$, resulting in a ferromagnetic coupling with the $S=2$ state of the Mn$^{+3}$ by Hund's rule. In contrast, in $N-1$ the ligand spin may have no overlap with the $e_{g}$ orbitals of Mn$^{4+}$ and it may be located near the $t_{\text{2g}}$ orbitals to which it couples anti-ferromagnetically, as observed in $N-1$.

The total spin of states $i$ and $iii$, as shown in Fig. 4, are 1 and 5/2, respectively, which amounts to a spin difference of 3/2. SET transitions between these ground states are thus rendered forbidden, as indicated by the red cross in Fig. 4. At high magnetic fields, the high-spin state $ii$ ($S=2$) becomes the ground state for $N-1$, lowering the ground-state spin difference to 1/2, whereby the GSSB is lifted.

This work shows that the act of charging a nano-scale object is not merely a consecutive filling of the lowest unoccupied orbitals, whereby the spin changes by $\pm$1/2. Rather, the act of charging can have a strong influence on the internal exchange couplings of the object, to a degree that ground-state transitions cannot be realized by single charge carriers. A thorough understanding of the resulting GSSB may therefore prove to be crucial in high-spin devices with applications in spintronics \cite{bogani2010molecular,sanvito2011molecular,clemente2012magnetic} and quantum computing \cite{leuenberger2001quantum,thiele2014electrically,shiddiq2016enhancing}.

%% SUMMARY
In summary, we have demonstrated GSSB in a single-molecule junction, which can be reversibly lifted by driving the system through a magnetic ground-state transition with an external magnetic field. The blockade results from a sign change of the exchange coupling upon charging, which causes the ground-state spin of subsequent charge states to differ by more than 1/2. This work demonstrates how the act of charging can induce a magnetic phase transition in a high-spin device by which resonant transport is completely suppressed.

\bigskip

\begin{acknowledgments}
This work was supported by the Netherlands Organisation for Scientific Research (NWO/OCW), as part of the Frontiers of Nanoscience program, and the ERC Advanced Grant agreement numbers 240299 (Mols@Mols) and 788822 (Mol-2D). P.G. acknowledges a Marie Skłodowska-Curie Individual Fellowship under grant TherSpinMol (ID: 748642) from the European Union's Horizon 2020 research and innovation programme. The work in Spain is supported by the Spanish MINECO (Unit os Excellence Maria de Maeztu MDM2015-0538 and Project MAT2017-89993-R co-financed by FEDER) and the Generalitat Valenciana (PROMETEO Programme). The Center for Quantum Devices is funded by the Danish National Research Foundation.
\end{acknowledgments}

\end{document}

\title{Supplementary Information --- Ground-state spin blockade in a single-molecule junction}

\author{J. de Bruijckere}
\affiliation{Kavli Institute of Nanoscience, Delft University of Technology, 2628 CJ Delft, The Netherlands}
\author{P. Gehring} 
\affiliation{Kavli Institute of Nanoscience, Delft University of Technology, 2628 CJ Delft, The Netherlands}
\author{M. Palacios-Corella}
\affiliation{Instituto de Ciencia Molecular (ICMol), Universidad de Valencia, Catedr{\'a}tico Jos{\'e} Beltr{\'a}n 2, Paterna, 46980, Spain}
\author{M. Clemente-Le{\'o}n}
\affiliation{Instituto de Ciencia Molecular (ICMol), Universidad de Valencia, Catedr{\'a}tico Jos{\'e} Beltr{\'a}n 2, Paterna, 46980, Spain}
\author{E. Coronado}
\affiliation{Instituto de Ciencia Molecular (ICMol), Universidad de Valencia, Catedr{\'a}tico Jos{\'e} Beltr{\'a}n 2, Paterna, 46980, Spain}
\author{J. Paaske}
\affiliation{Niels Bohr Institute, University of Copenhagen, DK-2100 Copenhagen, Denmark}
\affiliation{Center for Quantum Devices, Niels Bohr Institute, University of Copenhagen, DK-2100 Copenhagen, Denmark}
\author{P. Hedeg{\aa}rd}
\affiliation{Niels Bohr Institute, University of Copenhagen, DK-2100 Copenhagen, Denmark}
\author{H.S.J. van der Zant}
\email[E-mail: ]{H.S.J.vanderZant@tudelft.nl}
\affiliation{Kavli Institute of Nanoscience, Delft University of Technology, 2628 CJ Delft, The Netherlands}

\maketitle

\section{Molecule synthesis}
The pyridine-functionalized Mn(III) Anderson POM was prepared by modifying a previously described procedure \citep{allain2008hybrid}: [$\text{N}(\text{C}_{4}\text{H}_{9})_{4}]_{4}[\alpha\text{-Mo}_{8}\text{O}_{26}$] (0.5 g, 0.232 mmol), Mn(OAc)$_{3}$ (0.089 g, 0.350 mmol) and $(\text{HOCH}_{2})_{3}\text{CNHCO(4-C}_{5}\text{H}_{4}\text{N})$ (0.185 g, 0.818 mmol) were refluxed overnight under an argon atmosphere in 18 mL acetonitrile. The resulting solution was cooled to room temperature. Diethyl ether was added to obtain a precipitate, which was isolated by centrifugation and recrystallised in dimethylformamide.

\section{Device fabrication}
The transport device was created by depositing a bow-tie-shaped Au nanowire with a thickness of 12 nm, suitable for electromigration and self-breaking, on a pre-patterned gate electrode consisting of a 60 nm-thick Pd strip, covered with 5 nm of ALD-grown Al$_{2}$O$_{3}$. The nanowire was contacted to Au contact pads by 85 nm-thick Au connection lines. All Au and Pd structures were defined by standard e-beam lithography and e-beam evaporation techniques.

\section{Transport measurements}
All transport measurements shown in this work are performed in a dilution refrigerator with a base temperature of $T \approx 40$ mK. The d$I$/d$V$ maps in Fig. 1b, Fig. 2 and Fig. 3a are recorded in DC, whereas the magnetic field dependence in Fig. 3b is recorded using a lock-in amplifier, to achieve a higher signal-to-noise ratio. The magnetic field is applied with a vector magnet oriented out-of-plane with respect to the sample's substrate, except for the measurements shown in Fig. S1.

\section{Magnetic field dependence of the differential conductance spectrum in $N$}

\begin{figure}
\includegraphics{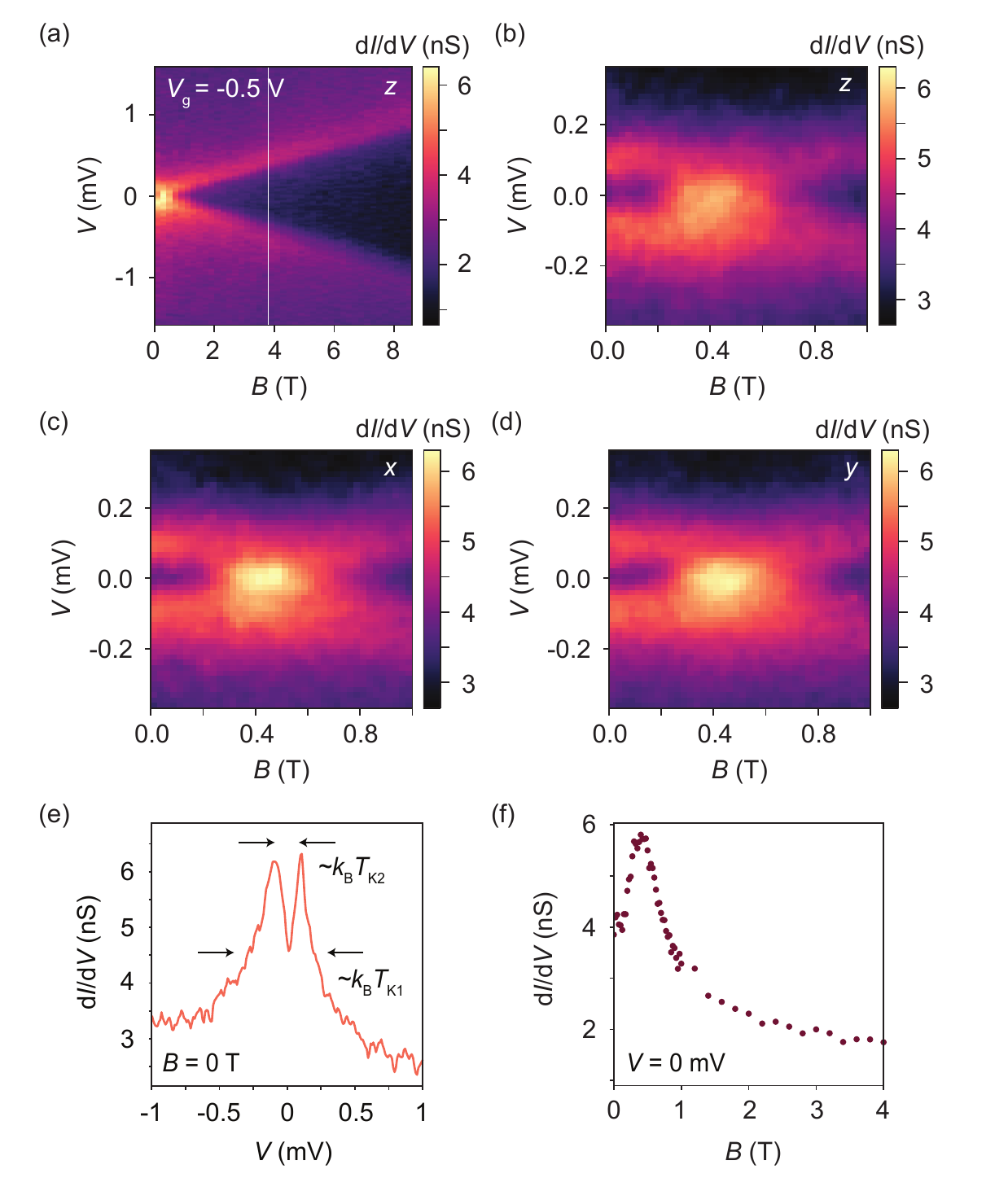}
\caption{Magnetic field dependence of the $\text{d}I/\text{d}V$ spectrum in $N$. All maps recorded at $V_{\text{g}}=-0.5$ V. (a) Magnetic field applied in the $z$-direction, i.e., perpendicular to the current, out of the substrate's plane. (b) Same $B$-field direction as in (a), but in a smaller $B$ and $V$ range. (c) Same as (b), with field applied along $x$, i.e., perpendicular to the current, in the substrate's plane. (d) Same as (b), with field applied along $y$, i.e., parallel to the current. (e) and (f) show the $\text{d}I/\text{d}V$ spectrum at zero magnetic field and the linear conductance in magnetic field, respectively.}
\label{fig:magbiasN}
\end{figure}

Figure S\ref{fig:magbiasN} shows four panels with the magnetic field dependence of the d$I$/d$V$ spectrum in charge state $N$ ($V_{\text{g}}=-0.5$ V). The data in Fig. S\ref{fig:magbiasN}a is recorded in a large magnetic field range, with the magnetic field directed out-of-plane ($z$-axis). One excitation splitting linearly with magnetic field can be identified. At low magnetic field, around 0.4 T, the excitation appears as a peak centered at zero bias. This peak is presumably due to a Kondo resonance that results from the spin degeneracy of the ground state. The expected COT excitations at finite magnetic field correspond to transitions from the $m=-5/2$ state of $iii$ ($S=5/2$) to the $m=-3/2$ state of both $iii$ ($S=5/2$) and $iv$ ($S=3/2$). Other transitions are forbidden due to the spin selection rules. Resolving these two excitations in the data is hindered by the Kondo peak and the small difference in excitation energy. 

A detailed map of the low-magnetic field range is shown in Fig. S\ref{fig:magbiasN}b, in which a dip can be observed below 0.4 T. The transition in magnetic field from a dip at 0 T to a peak at 0.4 T that splits again at higher fields, is independent of the direction in which the magnetic field is oriented, as can be seen in Fig. S\ref{fig:magbiasN}c and Fig. S\ref{fig:magbiasN}d. These maps are recorded with the magnetic field oriented in-plane, in two perpendicular directions. The peak-dip structure can be clearly observed in Fig. S\ref{fig:magbiasN}e where the d$I$/d$V$ spectrum at $B=0$ is presented. Figure S\ref{fig:magbiasN}f shows the linear conductance as a function of magnetic field, which is non-monotonic in the low magnetic field range.  

The independence of the magnetic field direction excludes the possibility of the dip being the result of a zero-field splitting by magnetic anisotropy, as for single-molecule magnets \cite{burzuri2015observing}. Similarly, a splitting due to Dzyaloshinskii-Moriya interaction depends strongly on the orientation of the magnetic field \cite{herzog2010dzyaloshinskii} and can be excluded. 

Instead, the data shows striking similarities with a two-stage Kondo effect \cite{van2002two}. This phenomenon can exclusively be observed in systems with $S > 1/2$, and involves two different Kondo energy scales: $k_{\text{B}}T_{\text{K1}}$ and $k_{\text{B}}T_{\text{K2}}$ ($T_{\text{K2}} < T_{\text{K1}}$). The former is associated with the first screening stage, whereby the spin is reduced by 1/2, yielding a peak of width $\sim k_{\text{B}}T_{\text{K1}}$; similar to the standard Kondo effect. The second stage reduces the spin by an additional amount of 1/2, which yields a zero-bias dip of width $k_{\text{B}}T_{\text{K2}}$. The dip results from the fact that the second screening stage quenches the first screening stage below $eV \sim k_{\text{B}}T_{\text{K2}}$ \cite{pustilnik2001kondo}. The spectrum in Fig. S\ref{fig:magbiasN}e contains both these features. The two Kondo energy scales are tentatively indicated by the arrows. As the magnetic field is increased, the second stage of Kondo screening stage will be suppressed, recovering the zero-bias peak of width $\sim k_{\text{B}}T_{\text{K1}}$: Here, at 0.4 T. At even higher fields, the Kondo peak associated with $T_{\text{K1}}$ splits in two, in a similar way as a standard spin-1/2 Kondo. The presence of a two-stage Kondo effect is also reflected by the non-monotonic evolution of the linear conductance in magnetic field \cite{van2002two,pustilnik2001kondo}, as shown in Fig. S\ref{fig:magbiasN}f for this device.

\section{Tunability of the parabolic ground-state transition line}

\begin{figure}
\includegraphics{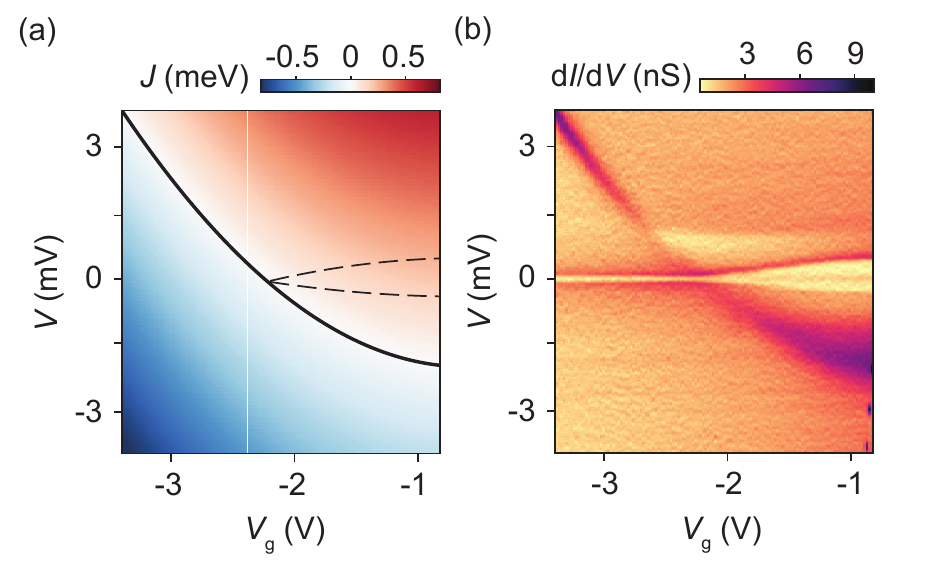}
\caption{(a) Value of $J$ according to equation \ref{eq:exchange}. The solid line indicates where $J=0$, which corresponds to a ground-state transition. The dashed lines indicate where $V=\pm 2J$, which is the condition for the low-energy COT excitations in this system. Parameters used: $\alpha = 0.2$, $\beta = 0.1$, $\gamma=-0.1$, $\delta=0.6$. (b) Experimental data in the same gate and bias voltage range as in (a).}
\label{fig:exchange}
\end{figure}

Here, we focus on the influence of the gate voltage and bias voltage on the exchange coupling between the high-spin Mn-center and the ligand spin in charge state $N-1$. We find that the exchange coupling is approximately given by the phenomenological expression
\begin{equation}
J=\alpha + \beta V + \gamma (V_{\text{g}} - \delta)^2,
\label{eq:exchange}
\end{equation}
in which $\alpha = 0.2$, $\beta = 0.1$, $\gamma=-0.1$, $\delta=0.6$. This quantity is plotted as a function of $V$ and $V_{\text{g}}$ in Fig. S\ref{fig:exchange}a, in the range of the experimental data shown in Fig. S\ref{fig:exchange}b. This region corresponds to the $N-1$ region of Fig. 2b. The solid line in Fig. S\ref{fig:exchange}a indicates where $J=0$, i.e., where the ground-state transition from $i$ to $ii$ takes place, as seen in the experimental data. The same expression for the exchange coupling is used to simulate the d$I$/d$V$ spectra in Fig. 3c. The dashed lines if Fig. S\ref{fig:exchange}a indicate where the COT excitations from $i$ to $ii$ are expected, according to the condition $V=\pm 2J$. The same gate and bias voltage dependence of these excitations is present in the experimental data shown in S\ref{fig:exchange}b.

\section{Magnetic field evolution of differential conductance maps}

\begin{figure}
\includegraphics{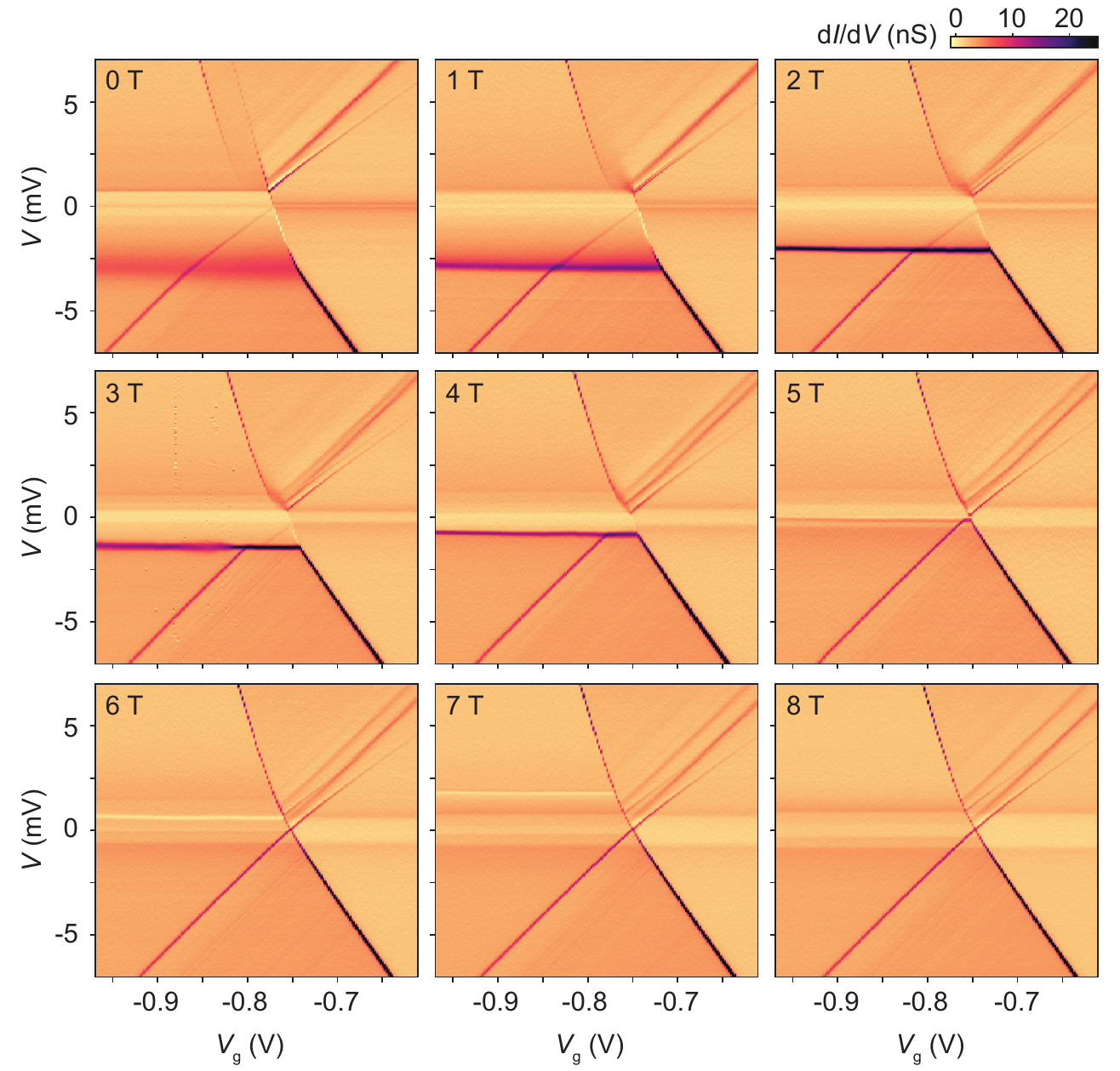}
\caption{Differential conductance maps at different magnetic fields. All maps are recorded in the same gate and bias voltage range at the indicated $B$-field. The GSSB is gradually lifted as the ground-state transition line moves upwards by the magnetic field.}
\label{fig:stabilities}
\end{figure}

A more detailed evolution of the lifting of the GSSB by the ground-state transition line is shown in Fig. S\ref{fig:stabilities}. The d$I$/d$V$ maps in this figure are all recorded in the same gate and bias voltage range, around the charge transition from $N-1$ to $N$, at different magnetic fields, as indicated in the panels. As the ground-state transition line crosses zero bias, between 5 T and 6 T, the charge degeneracy point is recovered, which indicates that the GSSB is lifted.

Figure S\ref{fig:stabilities} also shows that the ground-state transition line becomes narrower at higher magnetic field, i.e, when the spin multiplets are split. It is not fully understood how the magnetic field reduces the linewidth of the ground-state transition line.

%\bibliography{bibliography}